\documentclass[%
 aip,
 amsmath,amssymb,
 reprint,%
]{revtex4-2}

\usepackage{graphicx}
\usepackage{dcolumn}
\usepackage{bm}
\usepackage{color}

\newcommand{\changed}[1]{#1}

\begin{document}

\title{Invariant tori in dissipative hyperchaos}

\author{Jeremy P. Parker}
\email[Author to whom correspondence should be addressed: ]{jeremy.parker@epfl.ch}
\author{Tobias M. Schneider}
\affiliation{
Emergent Complexity in Physical Systems Laboratory (ECPS), \'Ecole Polytechnique F\'ed\'erale de Lausanne, CH-1015 Lausanne, Switzerland
}

\begin{abstract}

One approach to understand the chaotic dynamics of nonlinear dissipative systems is the study of non-chaotic yet dynamically unstable invariant solutions embedded in the system's chaotic attractor. The significance of zero-dimensional unstable fixed points and one-dimensional unstable periodic orbits capturing time-periodic dynamics is widely accepted for high-dimensional chaotic systems including fluid turbulence, while higher-dimensional invariant tori representing quasi-periodic dynamics have rarely been considered. We demonstrate that unstable 2-tori are generically embedded in the hyperchaotic attractor of a dissipative system of ordinary differential equations; that tori can be numerically identified via bifurcations of unstable periodic orbits and that their parametric continuation and characterization of stability properties is feasible.
As higher-dimensional tori are expected to be structurally unstable, 2-tori together with periodic orbits and equilibria form a complete set of relevant invariant solutions on which to base a dynamical description of chaos.
\end{abstract}

\maketitle

\begin{quotation}

Two-dimensional torus-shaped manifolds capturing non-chaotic quasi-periodic behaviour, and invariant under the dynamics, can be embedded within the state space of dynamical systems.
Such invariant tori have long been studied, for example in the transition to chaos, or as foliations of Hamiltonian systems. 
We show that invariant tori are also generically present in sufficiently chaotic dissipative systems, are structurally stable, and support the dynamics. Invariant tori thus provide a way of quantifying chaos in such systems. 
\end{quotation}

\section{Introduction}
Chaotic dynamics arise naturally from simple interactions in many physical systems ranging from fluid dynamics to electrical circuits and nonlinear optics. Studying the chaotic dynamics in terms of unstable non-chaotic \emph{invariant} solutions to the underlying evolution equations, which are embedded within a stable chaotic attractor, provides key insights into the observed physics. Two types of unstable invariant solutions are mostly studied: equilibria, zero-dimensional unstable fixed points in the system's state space; and periodic orbits, non-chaotic time-periodic solutions corresponding to one-dimensional loops in state space. Together with their entangled stable and unstable manifolds, equilibria form a skeleton of the chaotic attractor, while periodic orbits are often dense in the chaotic attractor and allow the computation of statistical averages of a chaotic dissipative out-of-equilibirum system. Indeed, the density of periodic orbits is sometimes taken as one of the criteria for a system to be called chaotic \citep{banks1992devaney}. 
Trajectories within chaotic attractors closely shadow unstable periodic orbits, so periodic orbits are often described as the `backbone' of chaos. 
For systems with dense periodic orbits, the construction of dynamical zeta functions \citep{bogomolny1992dynamical} allows statistical quantities to be evaluated as sums over the collection of all periodic orbits  \citep{cvitanovic1988invariant, eckhardt1994periodic, chandler2013invariant}.

While the importance of unstable zero-dimensional equilibria and one-dimensional periodic orbits is widely accepted, higher dimensional unstable invariant tori have never, to our knowledge, been considered as generic invariant structures embedded in a chaotic attractor and instead have only been studied as isolated exotic objects. Here we show that unstable tori are generically embedded in the chaotic attractor of a dissipative system and can be identified numerically. Moreover, unstable tori can be continued parametrically and their stability properties be computed. Consequently, including tori in a dynamical systems description of chaos now based on the complete set of invariant solution types appears feasible. 

In systems exhibiting a continuous symmetry, only a small number of periodic orbits are found, and instead `relative periodic orbits', a special case of 2-tori, take their role \citep{budanur2017relative}.
More generally, as the dimension of the chaotic attractor increases, we should expect to see higher-dimensional invariant structures. Spatiotemporal chaos, as is common in fluid dynamical situations, is associated with a large number of positive Lyapunov exponents (so-called `hyperchaos') and consequently a high dimensional attractor \citep{chian2010amplitude}, and invariant tori are thus presumed to be generic, and conjectured to be as important as the periodic orbits which have been extensively studied heretofore. 

For $M\geq 3$, in the absence of symmetries $M$-tori are structurally unstable \citep{newhouse1978occurrence,kaneko1986collapse}, and so we do not expect them to play an important role in chaotic dynamics, even for higher dimensional systems. 
In cases with a continuous symmetry, 2-tori become `relative' 2-tori, a special case of 3-tori which are structurally stable due to the symmetry, so become relevant, but these can be handled by careful modifications of the algorithms for 2-tori, analogously to the study of relative periodic orbits.
Therefore, the equilibria, periodic orbits and 2-tori represent a complete collection of non-chaotic solutions which form the skeleton of chaos.

In Hamiltonian dynamics, invariant tori have long been studied, playing a crucial role in KAM theory where they foliate phase space in integrable regions and act as boundaries between separate chaotic regions. In dissipative chaotic systems, by contrast, invariant tori (except invariant 1-tori, i.e. periodic orbits) have been largely neglected. The canonical example of such a system, the Lorenz system, on which much previous research in this area has concentrated \citep{eckhardt1994periodic,viswanath2003symbolic,maiocchi2022decomposing}, has only three dimensions, and so an invariant 2-torus would partition state space, and could not be embedded within the chaotic attractor.
Stable tori have nevertheless been studied in dissipative systems as the breakdown of a 2-torus is one of the possible routes to chaos \citep{ruelle1971nature, mackay1986transition, afraimovich1991invariant}.

Various authors \citep{kaas1987computation,dieci1995computation,moore1996computation,schilder2005continuation,haro2006parameterization,jorba2009computation,sanchez2010computation,sanchez2013parallel} have developed methods for the continuation of tori, and have successfully applied these to converge attracting tori and continue them as they become unstable, or to converge specific tori that arise from bifurcations.
\changed{These methods are now routine in Hamiltonian systems.}
However, to our knowledge, no previous work has attempted to find the generic unstable tori which are expected to be embedded within \changed{dissipative} hyperchaos.
\changed{
Generic invariant tori could allow the study of key phenomenology within certain dissipative chaotic systems, for which periodic orbits are seemingly rare. For example, in wall-bounded turbulence, it has proven difficult to find periodic orbit solutions which capture the interaction between different processes at different length-scales \citep{doohan2022state}. In the absence of phase locking, we expect the different temporal frequencies associated with these length-scales to lead to the dynamics manifesting as invariant tori rather than periodic orbits.
}

We  use the method of \citet{lan2006newton} for converging invariant tori with irrational rotation number, which they applied to converge a 2-torus in the Kuramoto-Sivashinsky partial differential equation. Though unstable in the full chaotic system, this torus is a stable attractor when the dynamics are restricted to the anti-symmetric subspace in which it was detected.
Our strategy for finding unstable 2-tori embedded in the chaotic attractor is based on one generic mechanism creating invariant tori, the bifurcation of an unstable periodic orbit. We will thus find unstable periodic orbits and continue these in parameter space until a complex-conjugate pair of Floquet multipliers loses or gains stability, and attempt to find the 2-torus that is born from the resulting Neimark-Sacker bifurcation.
For this to be possible, a minimum of five dimensions and three positive Lyapunov exponents is necessary, since an unstable periodic orbit which then undergoes a Neimark-Sacker bifurcation will have at the very least one real unstable Floquet multiplier, a complex conjugate pair, and one stable Floquet multiplier (to ensure a dissipative system), as well as the neutral direction.
It should be possible for invariant 2-tori to exist in four dimensional chaotic systems, but these cannot be continued from periodic orbits by Neimark-Sacker bifurcations, and thus it is not clear how they could arise.

\section{The model system}
We study, as a simple example, the 5D system proposed by \citet{vaidyanathan2014hyperchaos}, a modification of the Lorenz system:
\begin{align*}
    \dot x_1 &= a(x_2-x_1)+x_4+x_5,\\
    \dot x_2 &= cx_1-x_1x_3 -x_2,\\
    \dot x_3 &= x_1x_2-bx_3,\\
    \dot x_4 &= -x_1x_3+px_4,\\
    \dot x_5 &= qx_1.\\
\end{align*}
The system has one discrete symmetry, $(x_1,x_2,x_3,x_4,x_5)\mapsto(-x_1,-x_2,x_3,-x_4,-x_5)$ and, unlike in the standard Lorenz system, only one fixed point exists, at the origin.
With the choice of parameters $a=10$, $b=8/3$ and $c=28$--as is standard for the Lorenz system--and $p=1.3$, $q=2.5$, the chaotic attractor was found to have Lyapunov exponents $L_1 = 0.4195$, $L_2 = 0.2430$, $L_3 = 0.0145$, $L_4 = 0$, and $L_5 = -13.0405$. We will vary $p$ as a bifurcation parameter and keep all other parameters fixed.

\begin{figure}
    \centering
    \includegraphics[width=\columnwidth]{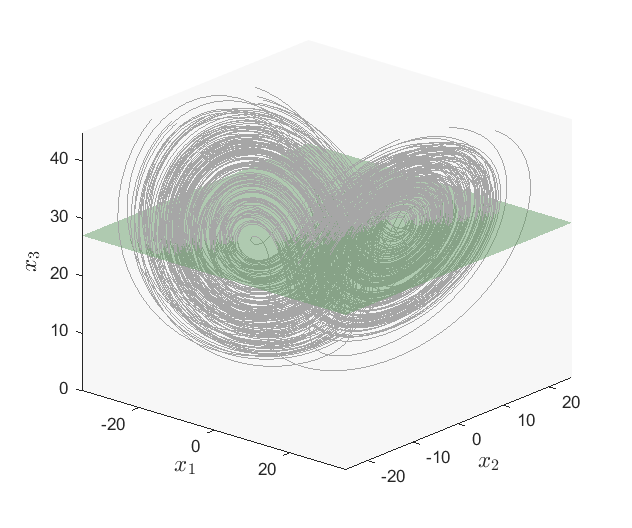}
    \caption{A typical trajectory of the system, at $p=1.3$, of length $T=50$, projected onto the first three variables. The Poincaré section employed, $x_3=27$, is also shown, here appearing as a 2-dimensional surface, though it is actually a four-dimensional sub-manifold.}
    \label{fig:poincare}
\end{figure}

We employ the customary Poincaré section for the Lorenz attractor, $x_3=c-1$, as shown in figure \ref{fig:poincare}. This yields a 4D subspace with which chaotic trajectories regularly intersect. Periodic orbits manifest as fixed points after a set number of iterations of the return map $\mathbf{F}(\mathbf{x})$ of this Poincaré section back to itself, and 2-tori manifest as invariant cycles, closed loops in the Poincaré section.

\section{Convergence of tori}
Below, we will assume for clarity of notation that only one iteration of the Poincaré return map is necessary, though in general this is not the case.
We follow the method of \citet{lan2006newton} which parameterises such a loop on the Poincaré section as $\mathbf{x}(s)\in \mathbb{R}^5$ with the cyclic variable $s\in[0,2\pi)$, so that $x_3(s)=c-1$ for all $s$.

For every point on the loop, we require that the return map rotates around the loop by an amount $\omega$, so that
\begin{equation}
\label{eq:torussol}
    \mathbf{x}(s+\omega)-\mathbf{F}(\mathbf{x}(s))=0.
\end{equation}
We assume the rotation number $\omega/2\pi$ to be constant around the loop, which is valid when it is irrational and thus Denjoy's theorem \citep{denjoy1932courbes} applies.
Equation (\ref{eq:torussol}) is solved using a Newton-like iteration for the fields $x_1(s), x_2(s), x_4(s)$ and $x_5(s)$, and the scalar quantity $\omega$. We solve
\begin{multline}
    \Delta\mathbf{x}(s+\omega)+\frac{d\mathbf{x}}{ds}(s+\omega)\Delta\omega-\mathbf{J}(\mathbf{x}(s)) \Delta\mathbf{x}(s) \\=\mathbf{F}(\mathbf{x}(s))-\mathbf{x}(s+\omega)
    \label{eq:tosolve}
\end{multline}
subject to the phase constraint
\begin{equation}
\label{eq:constraint}
    \int_0^{2\pi} \frac{d\mathbf{x}}{ds}(s)\cdot\Delta\mathbf{x}(s) \;ds = 0
\end{equation}
for $\mathbf{x}(s)$ and $\omega$
and update
\begin{equation}
    \mathbf{x}\gets\mathbf{x}+\epsilon\Delta\mathbf{x},\qquad \omega\gets\omega+\epsilon\Delta\omega.
\end{equation}
Here $\mathbf{J}(\mathbf{x})$ is the Jacobian of the Poincaré return map at a given point.

With $\epsilon=1$, this method is equivalent to the classical Newton-Raphson iteration and converges quadratically. However, in this case the algorithm will diverge unless a very good initial guess is available. Using a smaller value of $\epsilon$ increases the likelihood of convergence, at the cost of speed. We set the value of $\epsilon$ adaptively with the following procedure. At each step, the residual is compared with its value the previous step. If it has improved, we increase $\epsilon$ by a factor of $1.1$ and continue. Otherwise, we reset the state to its previous value, and halve the value of $\epsilon$. In this way, the algorithm is guaranteed to decrease the residual monotonically. However, it can still asymptotically approach a non-zero value, and thus is not guaranteed to converge.

The equation (\ref{eq:tosolve}) is particularly simple to solve in the case that $\mathbf{x}(s)$ is given by a finite Fourier series \citep{lan2006newton}.
A Fourier respresentation generally works well for $\mathbf{x}(s)$, but near an Arnold tongue (see section \ref{sec:arnoldtongues}), where the points on the loop cluster together, we see the effects of the Gibbs phenomenon. In this case, we switch to a piecewise linear interpolation.

Given a periodic orbit which intersects the Poincaré section at $\mathbf{x}_0$, near a Neimark-Sacker bifurcation such that the Jacobian of the return map has a complex conjugate pair of eigenvectors $\mathbf{e}$ and $\mathbf{e}^*$ with eigenvalues $re^{\pm i \theta}$ where $r\approx 1$, we construct an initial guess for the loop on the Poincaré section as
\begin{equation}
    \mathbf{x}(s)=\mathbf{x}_0+Re^{is}\mathbf{e}+Re^{-is}\mathbf{e}^*,
\end{equation}
where $R$ is a guess of the radius to be found by trial-and-error. An initial guess for the rotation number is given by $\omega=\theta$.

Once a torus has been converged at a given set of parameters, continuation is usually straightforward, by reconverging at new parameter values using the same algorithm, with the initial guess being given by constant, linear or quadratic extrapolation.

\section{Results}
\begin{figure}
    \centering
    \includegraphics[width=\columnwidth]{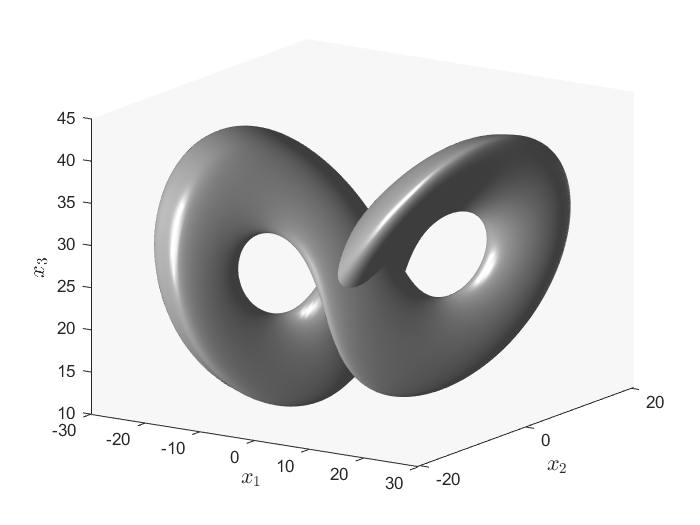}
    \caption{The torus $T_1$ at $p=0.7$. This three-dimensional projection of a two-dimensional manifold embedded in a five-dimensional space appears to self-intersect, but in reality the invariant torus does not. Note that this torus is invariant under the symmetry of the system.}
    \label{fig:T1}
\end{figure}

\begin{figure}
    \centering
    \includegraphics[width=\columnwidth]{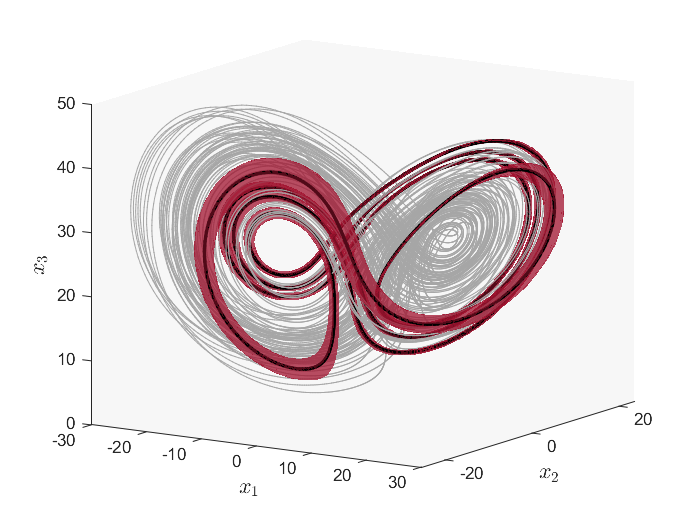}
    \caption{An unstable 2-torus $T_3$ (red), embedded within the chaotic attractor, together with the periodic orbit from which it arises (thick line) $P_3$ with period $T\approx5.51$, at $p=0.5033$. A trajectory of duration $200$ initialised very close to the torus is also shown (thin line), which briefly shadows the torus before being repelled away in the chaotic attractor. \changed{As with the previous figures, this is a projection showing only the variables $x_1$, $x_2$ and $x_3$.}}
    \label{fig:T3}
\end{figure}

\begin{figure}
    \centering
    \includegraphics[width=\columnwidth]{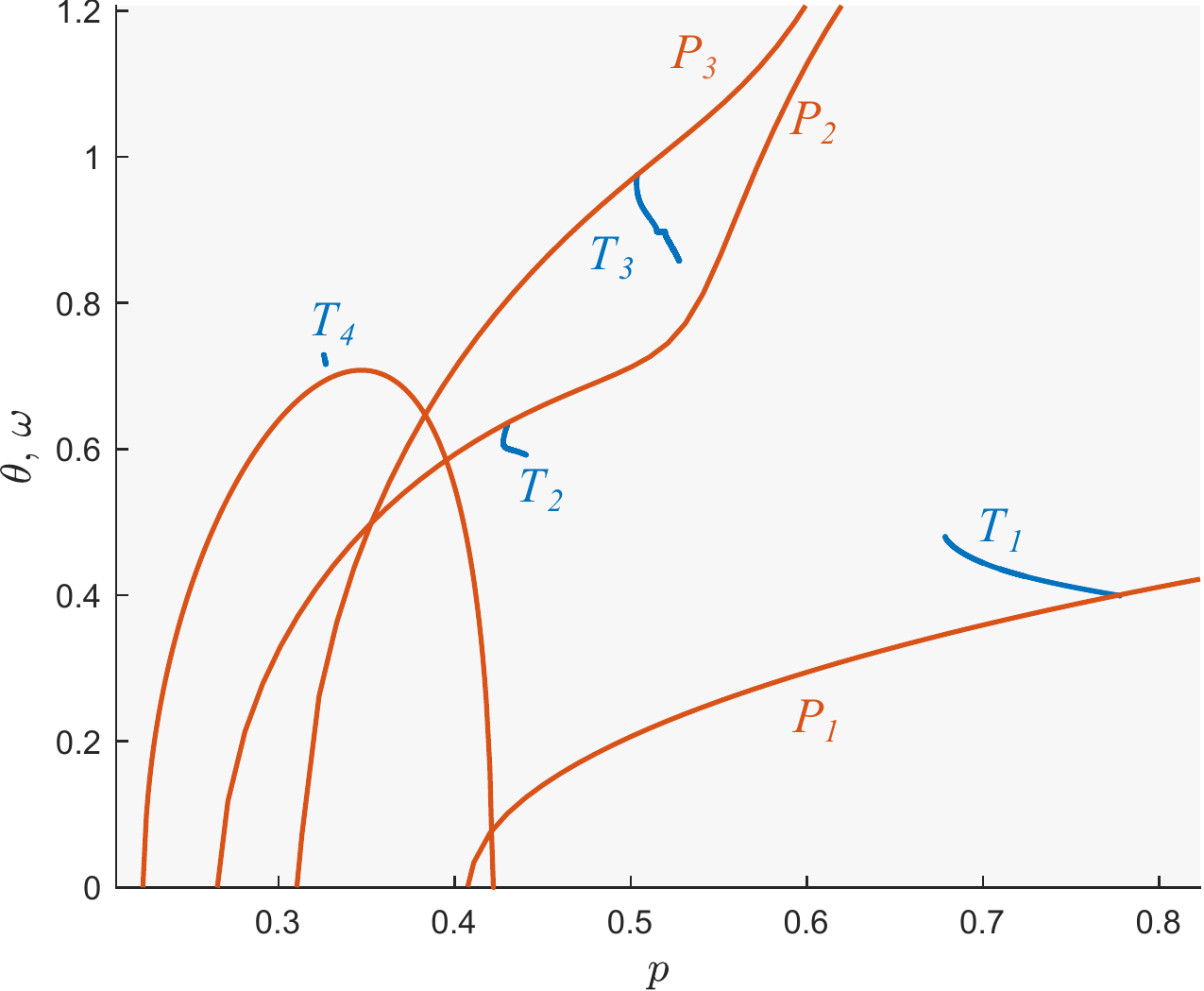}
    \caption{Bifurcation diagram showing all tori listed in table \ref{tab:table} and their associated periodic orbits.
    For the periodic orbits $P_n$ (red), the vertical axis shows the argument $\theta$ of the complex eigenvalue.
    For the tori $T_n$ (blue) which bifurcate from these orbits, the vertical axis shows the rotation number multiplied by $2\pi$, $\omega$.}
    \label{fig:all}
\end{figure}

A recurrent flow analysis was applied, at $p=1.3$, to find 8 unique unstable periodic orbits. Of these, 6 had an unstable conjugate pair of eigenvalues, which were then continued down in $p$ until a Neimark-Sacker bifurcation was found in 4 of the cases. In this way, four different invariant tori were converged, which are listed in the supplemental material. One such example which we call $T_3$ is shown in figure \ref{fig:T3}.
With a more robust recurrent flow analysis, we are confident that more, longer, periodic orbits could be found which would give rise to more tori. In this case, the computations would take much longer to converge, and indeed we already struggle when many reintersections of the Poincaré section are necessary.

Table \ref{tab:table} lists the tori we converged and their bifurcation points, and a diagram is given in figure \ref{fig:all}. Two of the four converged tori are invariant under the symmetry transformation $(x_1,x_2,x_3,x_4,x_5)\mapsto(-x_1,-x_2,x_3,-x_4,-x_5)$, including $T_1$, pictured in figure \ref{fig:T1}.
In all cases, the continuation eventually failed, either because of the presence of a phase locking (see section \ref{sec:arnoldtongues}) or simply because of the sensitive numerical precision required, resulting from the chaotic nature of the system.

\changed{In the case of $T_4$, the converged torus was found, by trial and error, some distance away from the bifurcation point. The numerical continuation of this, the most complicated of the tori which we successfully converged, failed before connecting back to the periodic orbit, hence the visible gap in figure \ref{fig:all}.}

\begin{table}
    \centering
    \begin{tabular}{c|c|c|c}
         Torus&NS bifurcation&\# intersections&Symmetric  \\ \hline 
         $T_1$& $p=0.776$& 2 &yes \\
         $T_2$& $p=0.430$& 6 &no \\
         $T_3$& $p=0.503$& 8 &no \\
         $T_4$& $p=0.327$& 8 &yes \\
    \end{tabular}
    \caption{The successfully converged tori, including the parameter value of the Neimark-Sacker bifurcation at which they appear, the number of intersections of the Poincaré section necessary to close the torus, and whether the torus is invariant under the discrete symmetry of the system.}
    \label{tab:table}
\end{table}

\subsection{Stability analysis}

\begin{figure}
    \centering
    \includegraphics[width=\columnwidth]{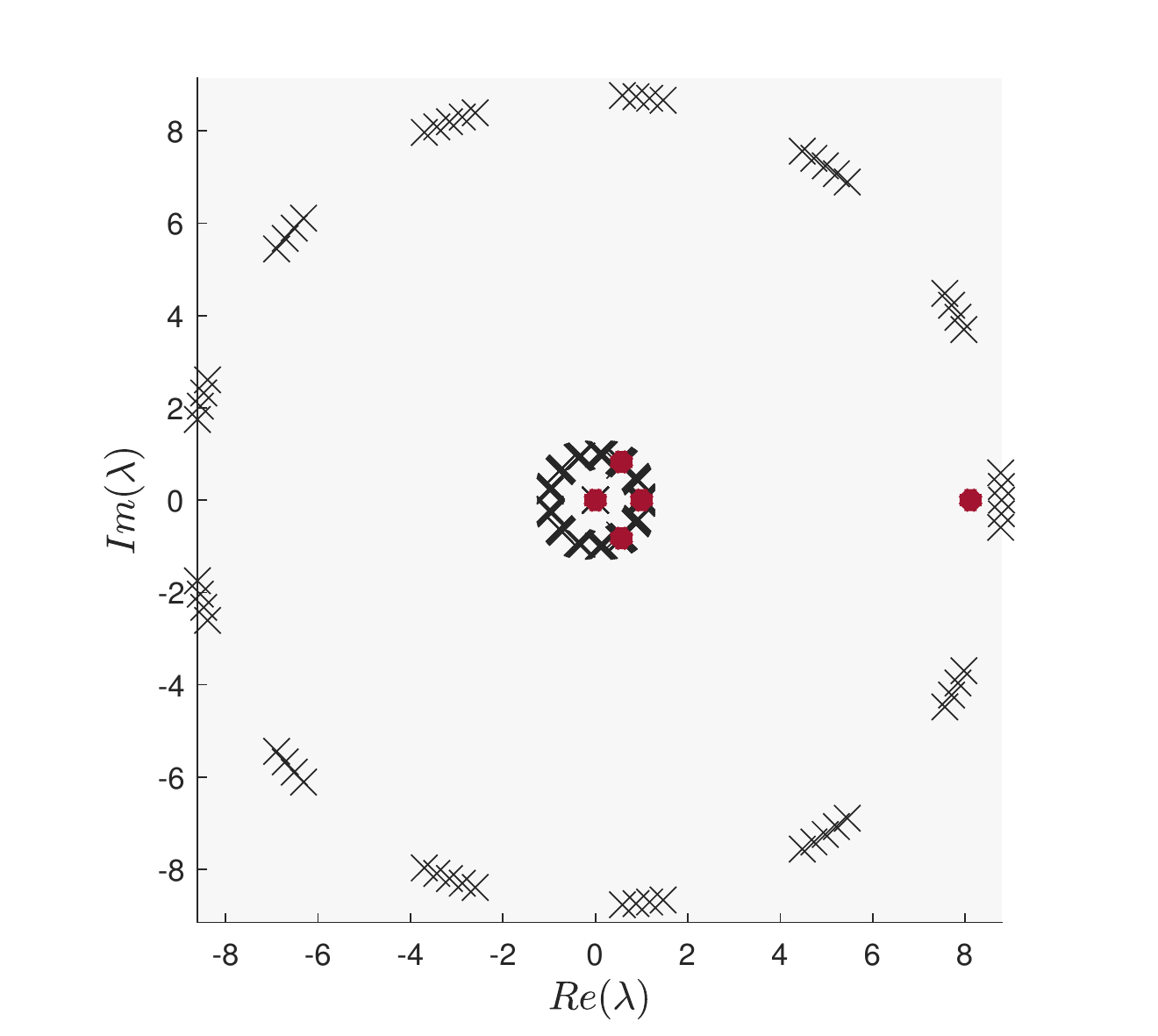}
    \caption{Eigenvalues of the Poincaré return map for the periodic orbit $P_3$ and the torus $T_3$ at $p=0.5033$.
    $P_3$ has four eigenvalues (red dots) in addition to the neutral direction along the orbit. Here we are close to the bifurcation point, so that the complex-conjugate pair have modulus slightly less than unity.
    The eigenvalues for $T_3$ should be smooth circles in the limit as $N\to\infty$ (see \citet{jorba2001numerical}). In addition to the neutral direction which manifests as a circle with radius 1, there are three circles: an unstable direction with radius about 9, a weakly stable direction with radius very slightly less than one (barely perceptible here) and a strongly contracting direction which appears to be a point at the origin.
    }
    \label{fig:eigenvalues}
\end{figure}

Following \citet{jorba2001numerical}, we measure the stability of the tori by calculating the eigenvalues of $\mathbf{T}_{-\omega}\mathbf{J}$, where 
$\mathbf{J}$ is the Jacobian of the return map applied to every point simultaneously,
and $\mathbf{T}_{-\omega}$ is the operator which rotates points around the loop by $-\omega$, to counteract the action of the return map. 
The set of eigenvalues is pruned for numerical accuracy to give circles whose radius, greater than or less than unity, determine how many stable and unstable directions the torus has. In total in this system there are four circles: one for each of the dimensions of the Poincaré section.
There is one circle with very small radius corresponding to the contracting direction, one with unit radius corresponding to a rotation around the loop, and two others, at least one of which must be unstable for a torus to be embedded within the chaotic attractor.
All four of the Neimark-Sacker bifurcations for which tori were converged were found to be subcritical, implying that the tori initially had fewer stable directions than the periodic orbits from which they were born. Figure \ref{fig:eigenvalues} shows the eigenvalues for $T_3$ and its associated periodic orbit, $P_3$, near bifurcation. For three of the four tori, a fold bifurcation was found close to the initial Neimark-Sacker bifurcation, so that the tori then became more stable, as in figure \ref{fig:T3bifurcation}.

\subsection{Continuation past Arnold tongues}
\label{sec:arnoldtongues}

\begin{figure}
    \centering
    \includegraphics[width=\columnwidth]{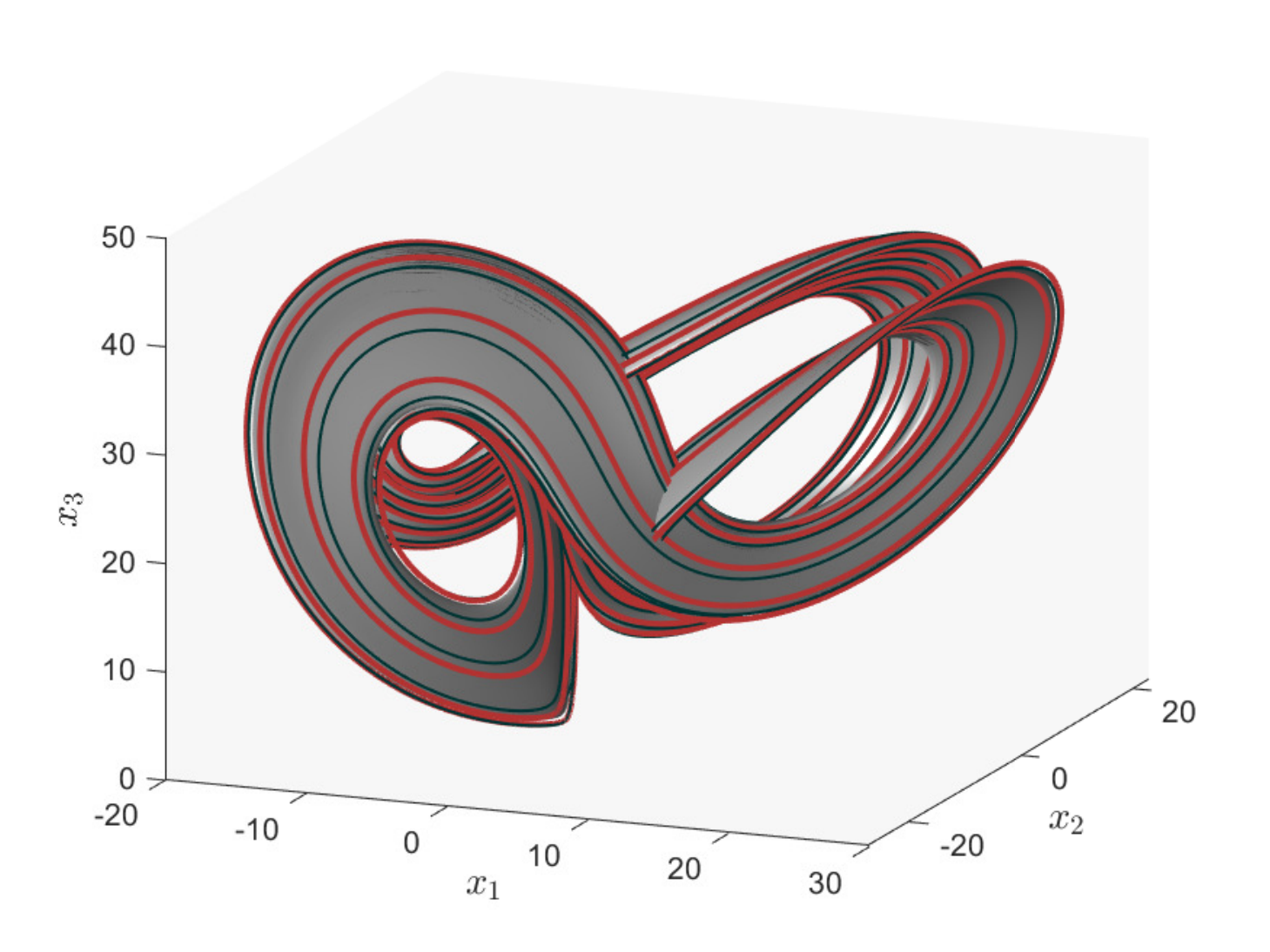}
    \caption{Periodic orbits on the torus $T_3$ at $p=0.5157$. The orbit shown with a thick red line is stable when the dynamics are restricted to the torus, and has period $T\approx40.28$, roughly 7 times longer than that of the original orbit $P_3$. The orbit shown with a thin black line is unstable, with $T\approx40.40$.
    The full torus at $p=0.51$ is also shown, to aid visualisation.}
    \label{fig:tongue}
\end{figure}

The rotation number $\omega/2\pi$ of a torus varies continuously as the parameters of the system are varied. Rational rotation numbers exist for finite regions of parameter space called Arnold tongues \citep{kuznetsov1998elements}. At any rational rotation number, the dynamical behaviour is quite different from the dense quasiperiodic orbit that exists on the torus at irrational rotation numbers. Instead, a phase locking is present and at least two periodic orbits--one attracting and one repelling--exist on the torus. In general, these periodic orbits will have extremely long periods, much longer than the period of the orbit from which the torus bifurcates, and the dynamics will be computationally indistinguishable from an irrational rotation number. Indeed, between any two steps with different rotation number in a numerical continuation of a torus, an infinite number of Arnold tongues have been crossed. However, for rotation numbers which are fractions with low denominators, such as $1/3$, $1/2$ and $1$, the phase locking is sufficiently strong that the Arnold tongues are wide regions of parameter space and the resulting periodic orbits have relatively short periods. In these cases, most algorithms for converging tori including that of \citet{lan2006newton} break down, as we can no longer appeal to Denjoy's theorem to enforce a loop parameterisation equivalent to a constant phase shift. Figure \ref{fig:tongue} shows the two periodic orbits which exist on the torus $T_3$, in the phase-locked region where $\omega=2\pi/7$.

To continue a torus past an Arnold tongue, special care has to be taken.
Suppose the rotation number is given by the irreducible fraction $p/q$ in the phase-locked region.
Then as the region is approached, the points on the discretised loop will tend to cluster around $q$ locations. Reducing the number of points $N$ to $q$, and replacing the constraint (\ref{eq:constraint}) with $\omega=2\pi p/q$, the stable periodic orbit on the torus can be converged directly--and continued round the fold bifurcations at the boundaries of the phase-locked region to additionally give the unstable periodic orbit.
It is then possible to reconverge the full torus on the other side of the tongue, though this requires some care.
Simply adding points between those describing the periodic orbit with some interpolation strategy does not work in general, since the chaotic nature of the system and the high dimensionality means the new points are unlikely to be close to the torus.
The procedure we follow is thus:
\begin{enumerate}
    \item Just past the phase-locked region, we converge as well as we are able using $q$ points with linear interpolation, and the usual constraint (\ref{eq:constraint}).
    \item We replace each of the $q$ points with two points very close together, and reconverge.
    \item We repeat this process until the resolution is sufficiently high that we can continue the full torus without artifacts.
\end{enumerate}
This leaves us with $N=2^k q$ points, for some integer $k$. We aim for around $N=64$, which was found to be sufficient. After continuation some distance away from the phase-locked region until the clustering of points is not pronounced, we switch from linear interpolation to a Fourier representation of the loop, which requires reconverging.

\begin{figure}
    \centering
    \includegraphics[width=\columnwidth]{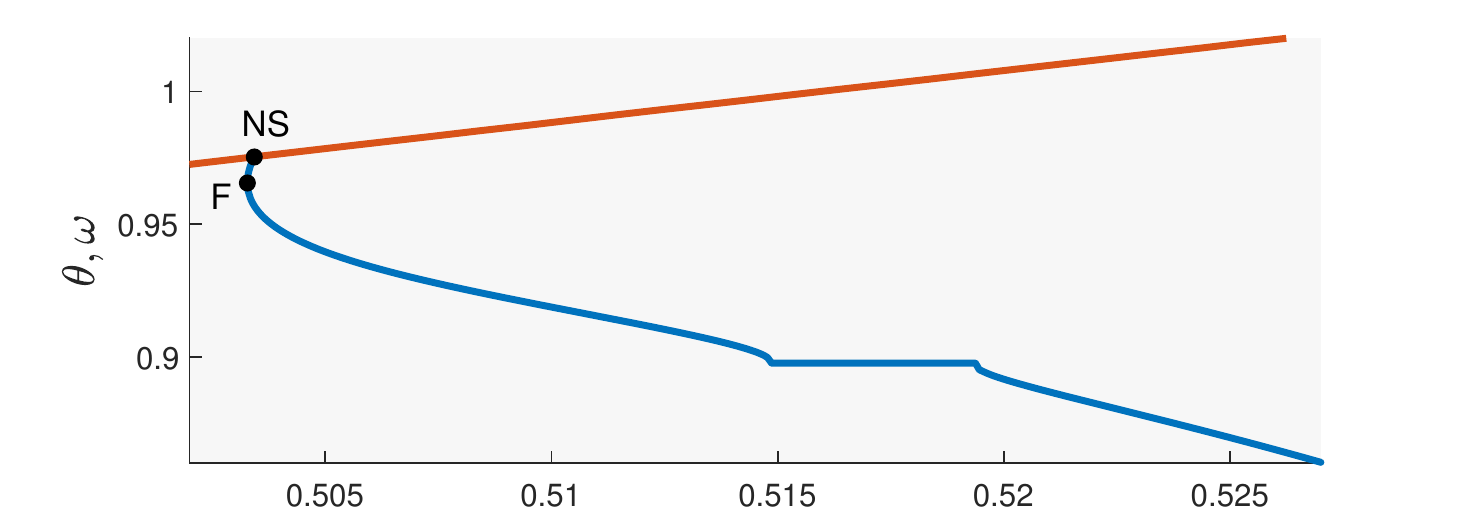}
    \includegraphics[width=\columnwidth]{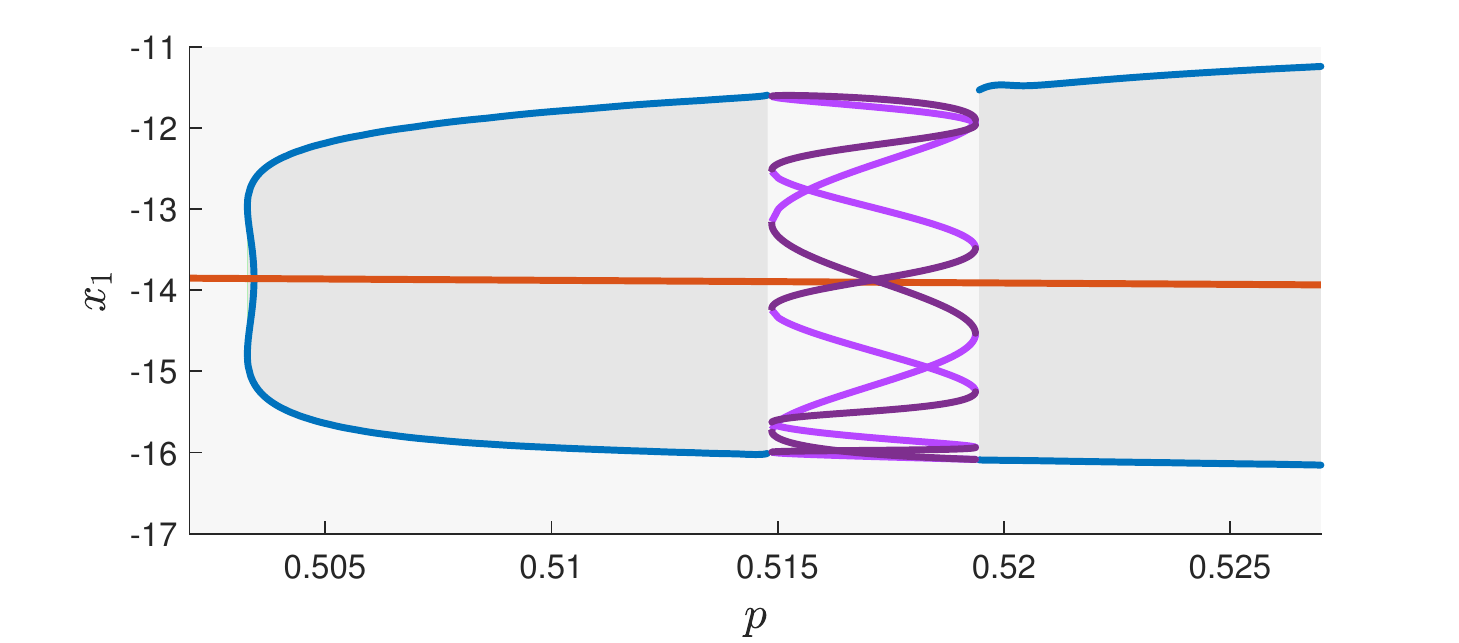}
    \caption{Bifurcation diagrams for the periodic orbit $P_3$ and the torus $T_3$ which arises from its Neimark-Sacker bifurcation (NS) at $p\approx5.034$, before undergoing a fold bifurcation (F). The upper figure shows a zoom of figure \ref{fig:all}. For $P_3$, the argument $\theta$ of the complex eigenvalue is plotted (red), and for $T_3$, the rotation number multiplied by $2\pi$, $\omega$ (blue).
    The lower figure shows the value of $x_1$ on the Poincaré section, which for $P_3$ is a single point and for $T_3$ is a range. Between $p\approx0.515$ and $p\approx0.519$, an Arnold tongue is encountered so that the rotation number is constant $\omega=2\pi/7$ and the full torus cannot be converged, but a pair of stable and unstable periodic orbits exists on the torus, each intersecting the Poincaré section seven times.}
    \label{fig:T3bifurcation}
\end{figure}

Figure \ref{fig:T3bifurcation} shows the continuation of $T_3$ from its Neimark-Sacker bifurcation, past a fold bifurcation and past an Arnold tongue with rotation number $1/7$.
Only $T_3$ was successfully continued past an Arnold tongue. In the case of $T_1$, two different phase lockings, with $\omega=2\pi/13$ and $\omega=6\pi/38$, were found at $p=0.675$, indicating that the invariant torus no-longer exists and continuation with a single parameter is impossible\citep{kuznetsov1998elements}.

\section{Conclusion}
We have demonstrated the generic existence of dynamically unstable invariant tori embedded within the chaotic attractor of a dissipative dynamical system. The 2-dimensional tori are identified by following Hopf-type bifurcations of unstable periodic orbits that are themselves embedded in the attractor. We moreover characterize the stability of the tori and continue them as parameters vary, including past an Arnold tongue. 
We have clearly demonstrated that such tori are structurally stable and exist in finite regions of parameter space. With sufficiently many long periodic orbits, we are confident that a large number of tori could be found at any parameter value.
As three- and higher-dimensional tori are in general structurally unstable, 2-tori together with previously studied equilibria and periodic orbits form a complete set of generic invariant solutions that a dynamical description of dissipative chaos should take into account. We expect tori to be especially significant when included in a generalized form of periodic orbit theory aimed at quantitatively describing statistical properties of a hyperchaotic system via expansions over invariant solutions. A 2-torus  intuitively captures more dynamical information than an individual periodic orbit, suggesting it would be associated a larger statistical weight and leading to more accurate expansions involving invariant solutions than a pure periodic orbit expansion. 

How exactly to incorporate 2-tori into periodic orbit expansions remains an open question. Ad-hoc formulae for the weights based on the unstable eigenvalues, like those suggested for periodic orbits \citep{zoldi1998spatially,kazantsev2001sensitivity,chandler2013invariant} are straightforward to invent, but a more rigorous derivation matching that of periodic orbits by \citet{ChaosBook} is not straightforward, since tori do not appear in the trace formulae derived from the assumption of dense periodic orbits.
One may alternatively attempt to consider the periodic orbits existing on the tori at arbitrarily close parameter values yielding rational rotation numbers. However, such periodic orbits will typically have very long periods rendering them practically indistinguishable from quasi-periodic dynamics but causing them to be discarded in a series truncation. Furthermore the periodic orbits exist only for small regions of parameter space and are not suitable for continuation.

Consequently, unstable invariant tori embedded in the chaotic attractor should be detected in physically relevant dissipative chaotic systems, though extending the studies to high-dimensional problems including fluid dynamical and other complex spatio-temporal systems will require significant advances in efficiency and robustness of methods for both finding guesses for the tori and for converging them.

This work was supported by the European Research Council (ERC) under the European Union's Horizon 2020 research and innovation programme (grant no. 865677).

\bibliography{references}

\end{document}